\documentclass[conference]{IEEEtran}
\IEEEoverridecommandlockouts
\usepackage{cite}
\usepackage{amsmath,amssymb,amsfonts}
\usepackage{tipa}
\usepackage{algorithmic}
\usepackage{graphicx}
\usepackage{cleveref}
\usepackage{threeparttable}
\usepackage{textcomp}
\usepackage{xcolor}
\def\BibTeX{{\rm B\kern-.05em{\sc i\kern-.025em b}\kern-.08em
    T\kern-.1667em\lower.7ex\hbox{E}\kern-.125emX}}
\begin{document}

\title{BanglaNum - A Public Dataset for Bengali Digit Recognition from Speech}

\author{\IEEEauthorblockN{Mir Sayeed Mohammad, Azizul Zahid, Md Asif Iqbal}
\IEEEauthorblockA{\textit{Department of Electrical and Electronic Engineering} \\
\textit{Bangladesh University of Engineering and Technology}\\
 }
}

\maketitle

\begin{abstract}
Automatic speech recognition (ASR) converts the human voice into readily understandable and categorized text or words. Although Bengali is one of the most widely spoken languages in the world, there have been very few studies on Bengali ASR, particularly on Bangladeshi-accented Bengali. In this study, audio recordings of spoken digits (0-9) from university students were used to create a Bengali speech digits dataset that may be employed to train artificial neural networks for voice-based digital input systems. This paper also compares the Bengali digit recognition accuracy of several Convolutional Neural Networks (CNNs) using spectrograms and shows that a test accuracy of 98.23\% is achievable using parameter-efficient models such as SqueezeNet on our dataset.\footnote{https://www.kaggle.com/datasets/mirsayeed/banglanum-bengali-number-recognition-from-voice}

\end{abstract}


\section{Introduction}
Speech recognition is an essential part of human-computer interaction that can facilitate convenient ways to work with intelligent machines. Nowadays, many computerized systems such as digital voice assistants, voice-based input systems, task-specific chat-bots on e-commerce sites or customer care services, personal assistants, etc benefit from speech recognition and can interact with the users naturally. These systems have been researched for a long time, particularly for languages like English, Chinese, etc. Modern speech recognition systems utilize artificial neural networks, and their performance is largely dependent on the availability and volume of accurately labeled natural language datasets. In the case of Bengali, speech datasets are less available which impedes the research and development of intelligent speech recognition systems that can understand Bengali. Instead of complex speech systems, digit recognition can simplify the task of building a voice-based automatic system. Thus to build a simple voice-based user interface using Bengali digits, we aimed to create a language dataset and develop a speech recognition system based on it. Consequently, we first collected speech data, preprocessed it, and then trained a neural network architecture for automatic speech recognition based on it. This paper is organized as follows - we first discuss the existing works on Bengali speech recognition. Then we describe our data collection and post-processing procedures in detail. Following that, we present experimental classification results using a convolutional neural network on the speech spectrogram of digit utterances from our dataset. Finally, we highlight the limitations of our work and discuss future research directions based on it.

\section{Related Work}

Research on speech recognition has been active since the 1930s and many studies have been published in the most widely spoken languages such as English\cite{pawar2014realization}, Chinese\cite{fu1996survey}, Indian\cite{kurian2014survey} and Portuguese\cite{4517917}. But studies to recognize Bengali speech have only begun in recent times. Here we have compiled the most up-to-date works on Bengali speech recognition problems that scholars have attempted to solve using various approaches. In \cite{sen4127691novel}, researchers created a Bengali Number Recognition (BNR) system based on Convolutional Neural Networks (CNNs). Their proposed algorithm correctly distinguishes Bengali spoken numbers (0-99) with an accuracy of 89.61\%. Besides, different Gaussian Mixture Model-Hidden Markov Models (GMM-HMM) have been explored in \cite{8554944} to develop a voice search module for the search engine "Pipilika". In \cite{9528088}, an alphabet of 39 phoneme symbols has been devised to categorize the data more precisely using Multi-layer Perceptron Classifier (MLPC) and Support Vector Machine (SVM). A BNR system from speech input using a CNN is built in \cite{9084030}. Here, a speech dataset consisting of 6000 utterances of 10 isolated Bengali digits has been introduced.
A Bengali isolated spoken numerals recognition system is created in \cite{10.1007/978-981-15-8752-8_9} that uses Mel Frequency Cepstrum Coefficients (MFCC) and GMM features. The work in \cite{10.1007/978-3-030-82269-9_29} suggests digit recognition from a mix of Bengali and English speech. In this work, the authors used an open-source dataset for English and created a new Bengali dataset in noisy environments from speakers of various ages, gender, and dialects. Then they used MFCC to extract features from the mixed dataset and a CNN classifier to train, test, and analyze the data. In \cite{10.1007/978-981-16-2543-5_8}, Short-Time Fourier Transform (STFT) is employed to create feature vectors in an HMM-based isolated BNR system. Also, a deep learning strategy for categorizing Bengali spoken digits is proposed in \cite{SHARMIN20201381}. It takes into account all aspects such as dialects, gender, and age groups. In our study, we propose a novel dataset for the recognition of Bengali digits from speech and follow the classification approach that was used to detect English language commands in the "Speech Commands" \cite{warden2018speech} dataset.

\section{Dataset Preparation}

The main purpose of this work is to introduce a dataset for detecting utterances of unit digits in the Bengali language that can be used in automated answering systems or voice-based digital assistants on e-commerce sites. In this section, we focus on the procedures followed during the preparation of this dataset.

\subsection{Data Collection}\label{ch:collection}
The data collection process was done inside a classroom environment, ensuring realistic ambient noise and minimal interference from other people talking. We have used MATLAB for recording the audio on a laptop using the microphone of generic headsets that people use for everyday work. This is also done to ensure that the dataset represents audio samples from everyday life and is easily deployable without requiring additional hardware. For generating the audio samples, native Bengali-speaking undergraduate student volunteers were asked to talk into the microphone for a duration of 40 seconds. During that time, the students had to cycle through the digits 0, 1, 2, 3, 4, 5, 6, 7, 8 and 9 for five times at varying rates, and then utter their student ID at a natural speed, all in Bengali. We have managed to collect audio samples from 40 people in total, with 35 male voice samples and 5 female voice samples. In accordance with the sample rate used in the work of P. Warden\cite{warden2018speech}, we have used a sampling rate of 16kHz and 24 bits of sample depth.


\begin{table}[ht]
\centering
\caption{Dataset Description}
\footnotesize
\label{table:dataset}
\resizebox{0.48\textwidth}{!}
{
\begin{tabular}{|p{0.035\textwidth}|p{0.02\textwidth}|p{0.02\textwidth}|p{0.02\textwidth}|p{0.02\textwidth}|p{0.02\textwidth}|p{0.02\textwidth}|p{0.02\textwidth}|p{0.02\textwidth}|p{0.02\textwidth}|p{0.02\textwidth}|p{0.03\textwidth}|}
\hline
\textbf{Digit} & 0   & 1   & 2   & 3   & 4   & 5   & 6   & 7   & 8   & 9   & \textbf{Total} \\ \hline
\textbf{Count} & 280 & 246 & 214 & 221 & 218 & 206 & 269 & 210 & 194 & 194 & \textbf{2252}  \\ \hline
\end{tabular}
}
\end{table}

\subsection{Data Processing}\label{ch:processing}
After collecting the audio samples, we had to convert them into a more usable format for training neural network models. Each utterance of a Bengali digit was cropped out from the full 40 second audio sample manually in MATLAB. Then all of these were brought to a uniform length of 8192 samples ($2^{13}$) per utterance by either cropping out extra data or by zero padding on both sides. This resulted in a dataset consisting of audio samples lasting 0.512 seconds each. After cutting out the audio samples, they were saved in the \textit{wav} format and sorted according to class labels into different folders as available in the dataset. The final dataset consists of a total of 2252 utterances of the 10 unit digits. \cref{table:dataset} summarizes the sample count in each class.

\section{Classification Experiments}

\begin{figure}[h]
    \centering
    \includegraphics[width=0.4\textwidth]{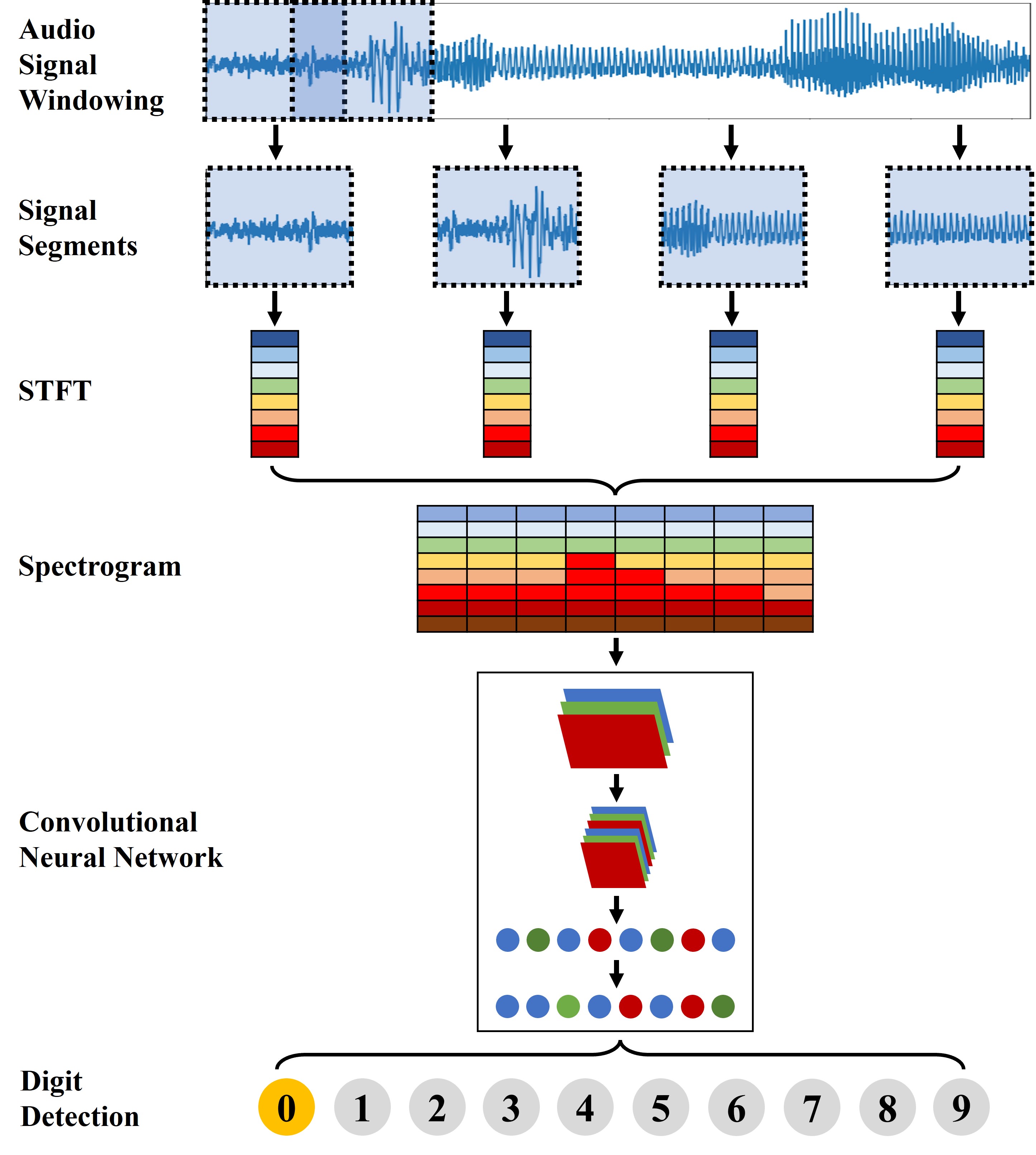}
    \caption{Overall system architecture.}
    \label{fig:algo}
\end{figure}

Experimental deployment of the dataset as a voice-based digital assistant required creating a speech command recognition model. For this purpose, we followed the method described in the work of P. Warden\cite{warden2018speech} where the voice spectrogram is fed through a simple convolutional neural network for classification. This section briefly describes the overall methodology shown in \cref{fig:algo}, as well as the comparative performances of different CNN architectures and the study for hyperparameter search.

\subsection{Calculating Spectrograms}\label{ch:spectrogram}

Audio spectrogram is a very efficient and effective way of representing the important features of speech signals. It performs exceptionally well in tandem with convolutional neural networks. For this, the signal is first segmented into chunks of small time lengths. Various window functions such as rectangular, Bartlett, Hamming, Hanning, etc can be used for this, and we have used the Hanning window given by \cref{eq:hann} in our experiments. Overlaps are introduced between windows to best capture the audio features. Then, for each of these segments, the Short Time Fourier Transformation (STFT) given in \cref{eq:stft} is evaluated and the absolute square value of the STFT is taken as given in \cref{eq:stft_sqr}, which represents the power in each frequency band. As a result, a 2-D array is formed in which the horizontal axis represents time, the vertical axis represents frequency, and the value of a point represents the power of the frequency component at any time. The overall process can be summarized by the following equations:

\begin{equation}\label{eq:hann}
    w[n] = 0.5(1-cos(2\pi \frac{n}{N})), 0\leq n \leq N, N = L-1.
\end{equation}
\begin{equation}\label{eq:stft}
    \textbf{STFT}\{x[n]\}(m, \omega) \equiv X(m, \omega)=\sum_{n=-\infty}^{\infty}x[n]w[n-m]e^{-i\omega n}
\end{equation}
\begin{equation}\label{eq:stft_sqr}
    \textrm{spectrogram}\{x[n]\}(m, \omega) \equiv |X(m, \omega)|^2
\end{equation}

Where w[n] is the window function, L is the window length, x[n] is the discrete time signal, X(m, $\omega$) is the STFT of the signal.

\begin{figure}[h]
    \centering
    \includegraphics[width=0.45\textwidth]{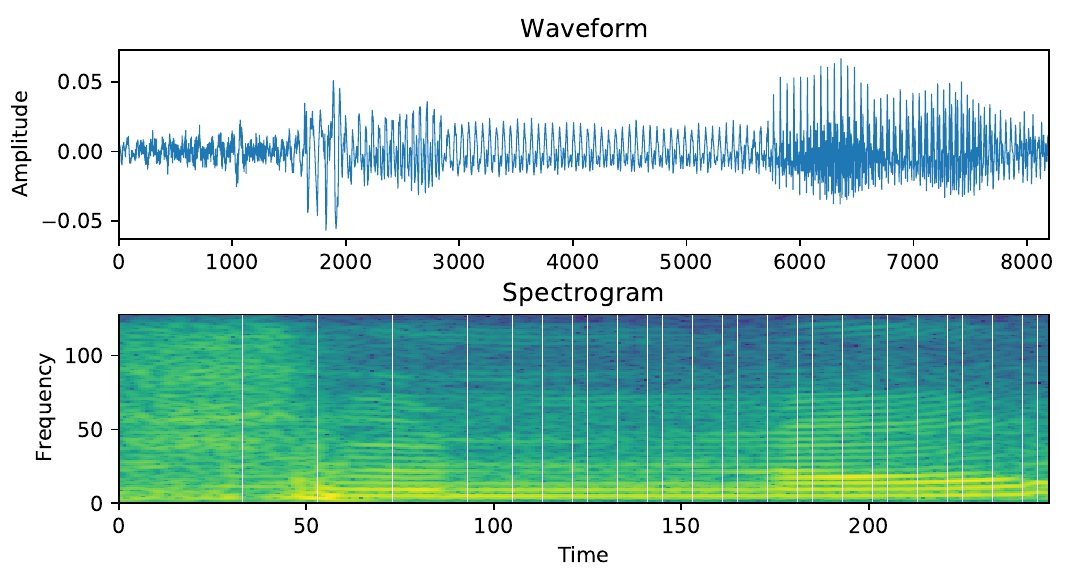}
    \caption{Example of an audio signal and its corresponding spectrogram.}
    \label{fig:spect}
\end{figure}

For our work, we have used a window size of 256 and a step size of 32 (overlap of 224 samples between windows). This gives a spectrogram with a dimension of $249\times129\times1$. Window and step sizes were determined by the hyperparameter search described in \cref{ch:hyperparameter}. \cref{fig:spect} shows the waveform and spectrogram of the utterance of \textit{0} in Bengali.

\subsection{Convolutional Neural Networks}\label{ch:models}

Spectrogram gives a visual representation of an utterance. This image can be used to classify the audio sample. For this classification task, we have experimented with different Convolutional Neural Network (CNN) architectures such as a basic custom CNN, SqueezeNet\cite{iandola2016squeezenet}, EfficientNet\cite{tan2019efficientnet} and ResNet50\cite{he2016deep}. Due to the small size of the dataset, we decided to use models that have a lower parameter count and do not 'remember' the training data entirely. The command recognition accuracy of these different models has been included in \cref{table:results}.


\subsection{Training}\label{ch:training}

\begin{figure}[h]
    \centering
         \includegraphics[scale=0.45]{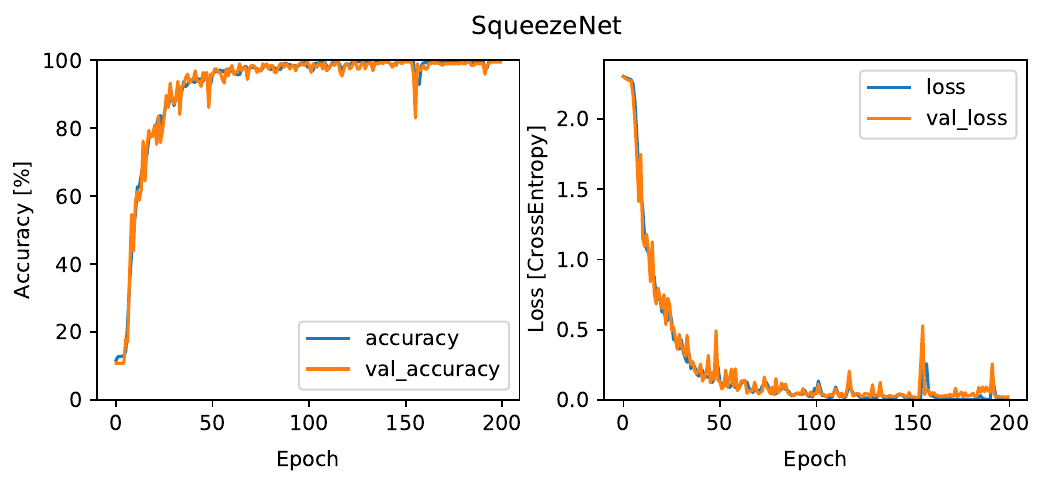}
    \caption{Accuracy and loss plots for SqueezeNet.}
    \label{fig:lossacc}
\end{figure}

For training the network, we have used 80\% of the dataset, and the other 20\% has been split evenly to be used as the validation and test sets. For optimization, we have used the Adam optimizer with a learning-rate of $10^{-4}$ and for the loss function, we have applied Sparse Categorical Cross-entropy. Each of the classification models was trained for 200 epochs with a batch size of 32. The batch size was determined by the hyperparameter search. During the training process, we monitored the training and validation losses as well as accuracy, so that we can see how our model improves over time in each successive epoch. \cref{fig:lossacc} shows the training and validation loss and accuracy.

\subsection{Results and Analysis}\label{ch:results}

We have tested several lightweight convolutional neural network models for training on the dataset and compared their performances. Lightweight models are preferable to avoid overfitting on the dataset as well as reduce inference times for real-time deployment. Less computational complexity enables the models to be implemented on CPU-only systems or single-board computers. So we have tested truncated versions of some of the state-of-the-art classification network architectures in our experiments. In \cref{table:results} we have reported the accuracies on the validation and test sets, as well as the parameter counts for each of the models.

\begin{table}[h]
\centering
\begin{threeparttable}
\caption{Bengali Number Classification Results}
\footnotesize
\label{table:results}
\begin{tabular}{|l|l|l|l|}
\hline
\textbf{CNN Network} & \textbf{Val Acc.} & \textbf{Test Acc.} & \textbf{Param. Count} \\ \hline
Basic CNN   & 0.9821	& 0.9513   & 61,984,525   \\ \hline
SqueezeNet  & \textcolor{red}{0.9955}  & \textcolor{red}{0.9823}             & \textcolor{red}{726,474}     \\ \hline
EfficientNetB0  & 0.9598    & 0.9646   & 4,061,801  \\ \hline
EfficientNetB0(Trunc)$^1$ & 0.9732  & 0.9602  & 1,356,877 \\ \hline
EfficientNetB1  & 0.9911   & 0.9735   & 6,587,469   \\ \hline
EfficientNetB1(Trunc) & 0.9821  & 0.9469 & 1,717,681 \\ \hline
EfficientNetB2  & 0.9732 & 0.9425 & 7,782,079   \\ \hline
EfficientNetB2(Trunc) & 0.9866 & 0.9558 & 1,967,653 \\ \hline
EfficientNetB3  & 0.9777	& 0.9602	& 10,798,181  \\ \hline
EfficientNetB3(Trunc)& 0.9911  & 0.9690  & 2,822,951 \\ \hline
EfficientNetB4 & 0.9732	& 0.9292	& 17,690,885 \\ \hline
EfficientNetB4(Trunc) & 0.9911 &	0.9602	& 4,400,205  \\ \hline
ResNet50  & 0.9911	& 0.9735	& 23,601,930    \\ \hline
\end{tabular}
\begin{tablenotes}
       \item [1] In the truncated models, the model head and the last convolutional layers are removed.
     \end{tablenotes}
    \end{threeparttable}
\end{table}

\begin{figure}
    \centering
    \includegraphics[width=0.4\textwidth]{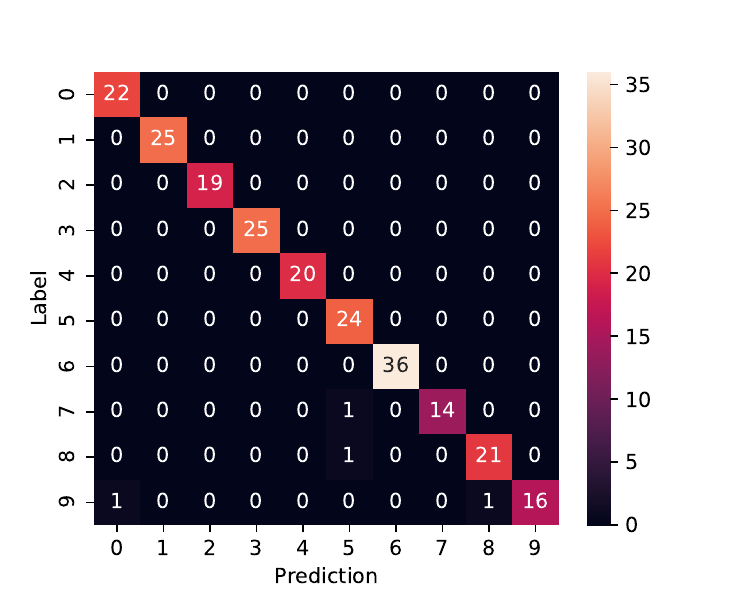}
    \caption{Confusion matrix for test set of SqueezeNet.}
    \label{fig:conf}
\end{figure}

In our tests, the SqueezeNet\cite{iandola2016squeezenet} architecture outperforms the other classification networks with a validation accuracy of 99.55\% and a test accuracy of 98.23\%. Some of the more complex architectures have lower performance compared to SqueezeNet despite having larger parameter counts. This implies that classifying Bengali digits from utterances can be accomplished using lightweight networks quite effectively. The confusion matrix of the test set is shown in \cref{fig:conf}.

We can see from the confusion matrix that among the 225 samples in the test set, digits 7 (phonetic transcription: /sh\begin{IPA}2\end{IPA}t/) and 8 (phonetic transcription: /\begin{IPA}2\end{IPA}t/) are misidentified by the network as 5 (phonetic transcription: /p\begin{IPA}2\end{IPA}ch/). This is because these numbers might sound similar when slightly mispronounced and can confuse even the native speakers. Additional processing might be required to correctly identify these two digits in Bengali speech.

\subsection{Hyperparameter Search}\label{ch:hyperparameter}

In order to select the correct batch size, spectrogram window length and step durations, we have conducted an extensive hyper-parameter search using the Basic CNN and SqueezeNet models. From these tests, the parameter values for which both models performed considerably well have been selected. We have several observations from the searching process as listed below:

\textbf{Batch Size:} Increasing the batch size had a detrimental effect on the model performances - the training curves showed instability along with a reduction of accuracy. From a set of 256, 128, 64, 32 - the best results were obtained for a batch size of 32. 

\textbf{Window Size:} the window size has a huge effect on the quality of information represented by the spectrogram. A large window can encapsulate a lot of frequency information but the time resolution is reduced. A smaller window loses frequency information but has better time resolution. Among window sizes of 1024, 512, 256, 128, 64 - the best result was obtained for the value of 256. This indicates that 256 samples offer optimum frequency and time resolutions at the sampling rate of 16 kHz. 

\textbf{Step Size:} for the tested step sizes of 256, 128, 64, 32, 16 - the best step size was found to be 32. A lower step size indicates that getting spectrograms at smaller intervals can prevent loss of information due to windowing.  


\section{Limitations and Scope for Future Work}

In our paper, we have collected audio samples in a classroom environment. So the potential sampling bias of this dataset could be reduced by collecting data from a more diverse group of people and varying environmental conditions. Moreover, the volume of the dataset could be increased for better generalization of learning algorithms. It could also be mixed with other Bengali speech datasets to augment the speech recognition capabilities of artificial neural networks. Finally, our reported results are mainly based on a classification algorithm based on spectrograms and CNNs - this could potentially be improved by other approaches such as wavelet transform based multi-resolution analysis, recurrent neural networks or transformer based architectures.

\section{Conclusion}

With the progress of technology, voice recognition based systems are being integrated into a variety of sectors of life. The Bengali voice recognition system is no exception. In this study, we have presented a novel dataset containing utterances of Bengali digits. Then we have described the data acquisition process and dataset details. Furthermore, we have shown the digit identification performance using a spectrogram and CNN based classification algorithm. Given that the experimental test accuracy is 98.23\% and might be further enhanced, we can infer that this dataset would be quite useful for developing Bengali voice based digital assistants for automated applications. 

\bibliographystyle{ieeetr}
\bibliography{citation}

\end{document}